# Structural phase transition and electronic structure evolution in $Ir_{1-x}Pt_xTe_2$ studied by scanning tunneling microscopy

Wei Ruan[1,*], Peizhe Tang[1,*], Aifang Fang[2], Peng Cai[1], Cun Ye[1], Xintong Li[1], Wenhui Duan[1,4], Nanling Wang[3,4], and Yayu Wang[1,4 †]

[1]*State Key Laboratory of Low Dimensional Quantum Physics, Department of Physics, Tsinghua University, Beijing 100084, P. R. China*

[2]*Institute of Physics, Chinese Academy of Sciences, Beijing 100190, P. R. China*

[3]*International Center for Quantum Materials, School of Physics, Peking University, Beijing 1 00871, China*

[4]*Collaborative Innovation Center of Quantum Matter, Beijing, P. R. China*

\* *These authors contributed equally to this work.*

[†] Email: yayuwang@tsinghua.edu.cn

**Abstract:** The $IrTe_2$ transition metal dichalcogenide undergoes a series of structural and electronic phase transitions when doped with Pt. The nature of each phase and the mechanism of the phase transitions have attracted much attention. In this paper, we report scanning tunneling microscopy and spectroscopy studies of Pt doped $IrTe_2$ with varied Pt contents. In pure $IrTe_2$, we find that the ground state has a 1/6 superstructure, and the electronic structure is inconsistent with Fermi surface nesting induced charge density wave order. Upon Pt doping, the crystal structure changes to a 1/5 superstructure and then to a quasi-periodic hexagonal phase. First principles calculations show that the superstructures and electronic structures are determined by the global chemical strain and local impurity states that can be tuned systematically by Pt doping.

**Introduction**

Transition metal dichalcogenides (TMDs) have been intensively investigated in the past few decades owing to their rich electronic phases such as charge density wave (CDW) [1], superconductivity [2], and topological insulating phase [3]. They are also expected to have potential applications in novel electronic and spintronic devices [4]. As one of the TMD materials, IrTe$_2$ exhibits a structural transition at temperature $T_s$ ~ 280 K [5-7]. A number of different mechanisms have been proposed to explain the origin of the structural phase transition, including Fermi surface instabilities induced by Ir 5$d$ orbital [6, 8] or van Hove singularity [9], and local bonding instabilities from Te 5$p$ orbital [5, 10, 11] or Ir dimerization [12-15]. Another intriguing phenomenon regarding this material is the emergent superconductivity with $T_c$ up to 3 K when doped or intercalated with Pt or Pd [5, 6, 16]. As shown in Fig. 1a, the superconducting (SC) phase boundary has a dome shape in proximity to an ordered state, very similar to that of the cuprates and iron pnictides high $T_c$ superconductors. It has become another model system for investigating the evolution from a possible density wave order to superconductivity.

Among the van der Waals coupled TMD compounds, IrTe$_2$ has a relatively strong interlayer coupling and the electronic structure is more three dimensional (3D). It has a lower ratio of $c/a$ = 1.372 ($c$ and $a$ are the inter and intralayer lattice constants respectively) compared to other layered CdI$_2$-type materials [5, 6, 11, 16, 17]. The $q$-vector of supermodulation has a $c$-axis component, significantly different from conventional in-plane CDW materials such as NbSe$_2$ [1]. Moreover, the extended 5$d$ electrons from Ir atoms are more likely to overlap with each other, thus having strong intralayer coupling. It has been found that upon Pt (Pd) doping, the lattice parameter $c$ gradually decreases and $a$ increases [16], effectively applying a chemical strain that reduces intralayer coupling but enhances interlayer coupling [16, 18-20]. It is likely that the structural phase transition and superconductivity in Ir$_{1-x}$Pt$_x$Te$_2$ are both related to intra and interlayer couplings, which can be tuned by the Pt concentration. Therefore, a thorough investigation of the doping dependence of electronic structures across the phase diagram is of great importance.

Due to its power of visualizing the lattice structure and electronic states at the atomic scale, scanning tunneling microscopy/spectroscopy (STM/STS) is an ideal probe for clarifying the mechanism of structural transition and its relationship with the electronic phases. Moreover, IrTe$_2$ can be easily cleaved due to its layered structure and exposes a charge neutral surface, which is ideal for STM investigations. In this paper, we report STM/STS studies on both pure and Pt doped IrTe$_2$ single crystals, which reveal a series of superstructures and electronic structures with varied Pt content. Theoretical calculations based on density functional theory (DFT) confirm that the structural phase transition is mainly controlled by the chemical strain and impurity states induced by Pt doping.

High quality IrTe$_2$ single crystals are grown by the self flux method [5]. STM experiments are performed with a low temperature ($T$) ultrahigh vacuum system. The IrTe$_2$ crystals are cleaved in the preparation chamber with pressure better than $10^{-10}$ mbar at 77 K, and then immediately transferred to the STM stage cooled at 5 K. An electrochemically etched tungsten tip is used for the STM measurements. Before each measurement the tip is treated and calibrated carefully as described elsewhere [21]. Topographic images are scanned in constant current mode and differential conductance (d$I$/d$V$) spectroscopy is measured by a lock-in amplifier with modulation frequency $f = 423$ Hz. All the measurements are performed at 5 K. Calculations are done in the framework of DFT using generalized gradient approximation (GGA) with the Perdew-Burke-Ernzerhof (PBE) functional [22] in Vienna *ab initio* Simulation Package (VASP) [23].

**Data description**

Figure 1d displays an 80 nm × 80 nm STM topographic image on the cleaved surface of undoped IrTe$_2$, showing a uniform supermodulation of parallel stripes. The inset is a zoomed-in image resolving the hexagonal lattice structure consisting of Te atoms as illustrated in Fig. 1a inset (the gray plane indicates the cleaving plane). In Fig. 1c we perform Fast Fourier Transformation (FFT), where the red dots denote the Bragg peaks, and there are 5 points between each pair of Bragg points along the (100) direction. This image clearly reveals the supermodulation wave vector $q = \frac{1}{6}(100)$ (denoted as the 1/6 phase), which can be directly compared with the schematic reciprocal lattice in Fig. 1b showing the 1/6 supermodulation. The vast majority of the surface we have explored exhibit long-range order of the 1/6 phase, confirming that it is indeed the ground state of undoped IrTe$_2$ [15, 24-26].

We further investigate the atomically resolved ground state superstructure and electronic structure of the 1/6 phase. As shown in the STM image in Fig. 2a, the atomic lattice structure of this phase consists of six different Te sites. Their appearance in the image is grouped into three pairs as marked by the colorful numbers in the figure, where a slight difference between atoms 3 and 6 demonstrates the 6 lattice constant periodicity. Fig. 2d shows the d$I$/d$V$ spectra, which is proportional to the local electronic density of states (DOS), on each Te site. In spite of the complex lineshape, the electronic structure exhibits two main features. (i) Each periodicity contains a 3+3' structure, where the second half is a direct translation of the first along the supermodulation direction. (ii) There are no gap-like spectroscopic features near the Fermi level ($E_F$). Instead, there are strong variations in the energy range of ±1 eV at different locations of the stripes. The observation of the structural distortion accompanied by the modification of d$I$/d$V$ spectra in large bias range strongly supports the idea that the deformation of Ir 5$d$ electronic state rather than the Fermi surface instability is the driving force for this phase transition [5].

The observation of 1/6 phase seems to be inconsistent with various bulk measurements such as electron and X-ray diffractions at relatively high $T$ [6, 10-12, 14], which instead suggest a 1/5 phase with modulation vector $q' = \frac{1}{5}(100)$ when projecting the bulk value $q'_B = \frac{1}{5}(101)$ onto the (001) surface. However, we do see, very occasionally, the mixed (3n+2) phase [25, 27-29] where 3$a$ and 5$a$ stripes distribute randomly, as shown in Fig. 3a. This is possibly induced by the remained unreleased strain or defect pinning, when the lattice evolves from the high $T$ 1/5 supermodulation to the ground state 1/6 phase, which is indeed visualized in high $T$ STM images [25]. In addition, we also find some small areas (~ 5 nm) of 1/5 phase shown in Fig. 3b, separated by abrupt slip of phases. We carefully examine the electronic structure away from these phase slip boundaries, and the averaged spectrum is shown as red curve in Fig. 4b.

We next turn to the Pt doped IrTe$_2$ with chemical formula Ir$_{0.97}$Pt$_{0.03}$Te$_2$ (PT3), which shows a structural phase transition at $T_s$ ~ 150 K and a SC transition at $T_c$ ~ 3 K. Fig. 4a displays an 80 nm × 80 nm area with 1/5 phase on PT3, which is more obvious in the atomically resolved image and FFT image shown in Fig. 4b and 4c. This phase differs from the 1/5 phase observed in parent compound in that it is actually long-range ordered, thus providing an ideal platform for us to study the 1/5 phase. However, the spatially averaged d$I$/d$V$ spectra of both phases shown together in Fig. 4d are highly similar to each other. We again hardly see any gap feature near $E_F$ but instead a hump at small negative bias, which corroborates our conclusion that the structural transition is not of Fermi surface instability type. The energy positions of features in these two curves agree very well, indicating little band structure reconstruction or rigid band shift as we dope Pt.

With further increase of Pt doping to Ir$_{0.95}$Pt$_{0.05}$Te$_2$ (PT5), the structural transition is totally suppressed and the SC transition $T_c$ is slightly lowered to ~ 2.8 K. The STM images in Fig. 5a and inset show that the stripe-like supermodulations are completely suppressed as expected from the phase diagram, but instead a 3-fold quasi-periodic pattern emerges as shown in the real space and Fourier transformed images. Therefore, the system evolves from the 1/6 phase to the 1/5 phase and then to the 3-fold quasi-periodic phase as we increase the Pt concentration. The spatially averaged d$I$/d$V$ spectrum shown as the red curve in Fig. 5b is almost featureless except for a zero bias peak (ZBP) and a peak at -300 mV. Compared to the spectra of the distorted phase, our data is consistent with the suppression of zero bias spectral weight upon structural transition.

We then examine the local Pt doping effects by looking at Figure 5e, where the perfect atomic lattice even persisting to the dark spots (marked by red dots) on a PT5 sample indicates that the dark spots are the Pt substitution sites underneath the surface Te layer. We count the number of dark spots on both samples, and the values are 1.3% for PT3 and 3.7% for PT5, which are roughly consistent with the nominal doping levels. Topographic images are acquired using different bias voltages from 2.0 V to -2.0 V, as shown in Fig. 5c-g. As can be seen clearly, the local C3 quasi-periodic structure spreading from each Pt substitution site exhibits strong variation with energy. The network formed by the spatially localized Pt impurity states is further corroborated by the d$I$/d$V$ maps shown in Fig. 6c-h, which directly map out spatial distribution of electron DOS at different energies. The spectra at different sites as marked by the colorful dots and ovals are plotted in Fig. 6a, which show strong spatial variation. We notice that the blue curve has the most pronounced ZBP near the center of each hexagonal islands, which is gradually suppressed from the island center to the Pt dark spots.

In order to better understand the mechanism of the structural transitions and the associated electronic states, we perform first principles calculations on each doping. For pure IrTe$_2$, DFT calculations with GGA-PBE functional typically underestimates the Ir-Ir in-plane bonding that mainly contributes to the intralayer coupling [29], and equally overestimate the interlayer coupling between IrTe$_2$ layers. So a slab model with a single layer of IrTe$_2$ is used to simulate the exposed surface in STM experiments. After full relaxation, it is found that the 1/6 phase can be stabilized without specifically setting the "best" Te positions [29], and its total energy is similar to that of the 1/5 phase from slab calculations. The calculated structure of 1/6 phase and local density of states (LDOS) are both compared with the measured STM image and d$I$/d$V$ as displayed in Fig. 2, and they show good agreement.

For Pt doped samples, we compare the measured d$I$/d$V$ spectrum on PT5 to the electronic DOS calculated based on the high $T$ undistorted phase of undoped IrTe$_2$ (Fig. 5b). The overall lineshape and peak positions show strong similarity. This is analogous to the fact that the electronic structure of the 1/5 phase in undoped sample and PT3 are also very similar to each other. Based on these observations, we conclude that the electronic structures exhibit no rigid shift as expected from charge carrier doping, and are insensitive to the Pt doping level as long as they are in the same structural phase. Thus, the doping of Pt mainly applies chemical strain to the sample and induces localized impurity states instead of global charge doping. Moreover, the integrated projected DOS (PDOS) at each atomic site (Fig. 6b) obtained by DFT calculations shows that the origin of the ZBP is Te 5$p$ electronic states, which is pronounced in the undistorted region near the center island, but gradually suppressed as entering the Pt dark spots where local lattice distortion occurs.

**Discussion**

Our STM results on pure IrTe$_2$ demonstrate that the ground state is the 1/6 phase, and large bias d$I$/d$V$ spectra supports the idea that the deformation of Ir 5$d$ electronic state rather than the Fermi surface instability is the driving force for this phase transition. The excellent agreement between the experimental data and theoretical simulation using slab model in this phase highlights the crucial role of intralayer Ir-Ir dimerization regarding the structural transition [12-15]. When the interlayer coupling is ignored, these dimers are arranged in the dense 1/6 pattern in the ground state. We notice that the dimerization is accompanied by pushing nearby Te atoms away from the Ir plane, and the displacement ~ 15% is comparable to the bulk scattering experiments [10, 12-14]. When adjacent layers are reintroduced, the shortened interlayer Te-Te distance as a result of Ir dimerization would induce strong interactions between dimers in adjacent layers, making each dimer offset by one atom with respect to dimers in both adjacent layers. Therefore the superstructure gains an out-of-plane component along the (101) direction. Comparing the electronic band structure in both distorted phase (Fig. 2d-e, Fig. 4d) and that calculated in undistorted phase (Fig. 5b), we find that Te bond shortening is accompanied by the suppression of LDOS around $E_F$. This is in agreement with optical and neutron scattering measurements [5, 24], which could be the effect of splitting of Te bonding and anti-bonding states.

We next discuss the effect of Pt doping on the structural transition and electronic structure. We notice that Pt doping barely inject charge carriers or change the overall d$I$/d$V$ lineshape, and the charged impurity states are rather tightly bounded to the Pt sites. Therefore at small Pt concentrations, the induced chemical strain is the main effect, which compresses the out-of-plane and elongates the in-plane lattices [10, 16]. The Ir-Ir intralayer bonding is then weakened, leading to a decrease in the energy gain from Ir 5$d$ electronic structures, which competes with the energy cost by deforming the lattice locally. As a consequence, the density of Ir dimerization gets reduced upon Pt doping, which changes the ground state from the 1/6 to the 1/5 phase, and finally to the phase without dimerization. Similar effects are observed in Se doped samples [11], static pressure effect [30], other transition metal substitutions [31], and variable temperature STM [25]. Though the thickness of each layer decreases (by less than 1%) with increasing Pt concentration, the shortened Te-Te interlayer bond length actually increases due to the suppression of associated Ir dimer. Therefore the system gradually recovers the polymeric Te chains across layers, which lowers the Te $5p$ splitting to give larger DOS at $E_F$, and makes the system more three dimensional. Both consequences are beneficial to the formation of a SC ground state.

Compared to the Ir ions, one additional $5d$ electron is bounded to each Pt site because no itinerant charge is injected. This localized electronic state and associated lattice distortion get more prominent with increasing Pt doping, which are visualized in the varied bias topographic images and d$I$/d$V$ maps (Fig. 5, 6) on sample PT5. More specifically, at the 5% doping level these states are dense enough to self-assemble into a quasi periodic network. The localized states start to stretch out by a few lattice constants and each has significant overlap with the nearest three ones. Compared to the Ir dimerization, this kind of distortion is more energetically favorable, which helps suppress the structural transition. However, the origin of such striking pattern is still an open issue and the possible candidates could be the formation of impurity band or orbital-ordered state.

In summary, our STM/STS results combined with DFT calculations reveal that Ir dimerization is the main driving force for the structural transition in IrTe$_2$. This dimerization could induce the interlayer Te-Te depolymerization, which greatly lowers the spectral weight at $E_F$. The Pt dopants in these materials mainly serves as a source of chemical strain as well as local electronic and lattice perturbations, which cause the reduction of Ir dimer density and Te depolymerization. As a result, the DOS at $E_F$ is enhanced and the system becomes more three dimensional, making the SC ground state more favorable.

**Acknowledgement**

This work was supported by the National Natural Science Foundation and MOST of China (grant No. 2011CB921901, 2011CB921701, 2012CB821403, and 2015CB921000).

**Figure Captions:**

FIG. 1. (a) Schematic phase diagram of $Ir_{1-x}Pt_xTe_2$. The inset displays the lattice structure, where the gray plane between two layers indicates the cleaving plane. (b) Schematic reciprocal lattice showing the in-plane 1/6 supermodulation marked by blue points. (c) Typical Fourier transformed STM image of $IrTe_2$. Bragg points are marked red in both (b) and (c). (d) Topographic image ($I = 50$ pA, $V = 1$ V) showing the pure 1/6 supermodulation of $IrTe_2$. The inset is a zoomed-in image with atomic resolution ($I = 0.3$ nA, $V = -20$ mV).

FIG. 2. (a) Atomically resolved image of the 1/6 phase ($I = 0.3$ nA, $V = -20$ mV), where the red dashed polygon highlights an area that can be directly compared with the lattice distortion in (b) calculated using the slab model. (c) Side view of the calculated structure. (d) d$I$/d$V$ spectra at different atomic sites as marked by the colorful numbers. (e) Calculated local density of states at each atomic site.

FIG. 3. (a) Mixed (3n+2) phase ($I = 50$ pA, $V = 0.5$ V) on $IrTe_2$ consisting of $3a$ and $5a$ stripes. (b) Short-ranged 1/5 phase ($I = 40$ pA, $V = 1$ V) on pure $IrTe_2$.

FIG. 4. (a) Topographic image ($I = 50$ pA, $V = 1$ V) showing the pure 1/5 phase of $Ir_{0.97}Pt_{0.03}Te_2$). (b) Atomically resolved image ($I = 0.2$ nA, $V = 20$ mV), and (c) Fourier transformed image showing the 1/5 peak. (d) Comparison of d$I$/d$V$ spectra of PT3 and the minority 1/5 phase on pure $IrTe_2$.

FIG. 5. (a) Topographic image ($I = 0.2$ nA, $V = 0.3$ V) showing the hexagonal quasi periodic pattern of $Ir_{0.95}Pt_{0.05}Te_2$. The inset is the Fourier transformed image showing the quasi super modulation wave vector. (b) Comparison of d$I$/d$V$ spectrum of PT5 and calculated LDOS of parent $IrTe_2$ in the undistorted phase. (c)-(g) Topographic images at different bias voltages of the same area on PT5 showing the quasi periodic pattern.

FIG. 6. (a) d$I$/d$V$ spectra at different sites on PT5. The inset shows the places where the spectra are taken. (b) Calculated total LDOS and PDOS summarized at Ir and Te atoms of the undistorted phase of pure $IrTe_2$. (c)-(h) d$I$/d$V$ maps at different bias voltages of PT5.

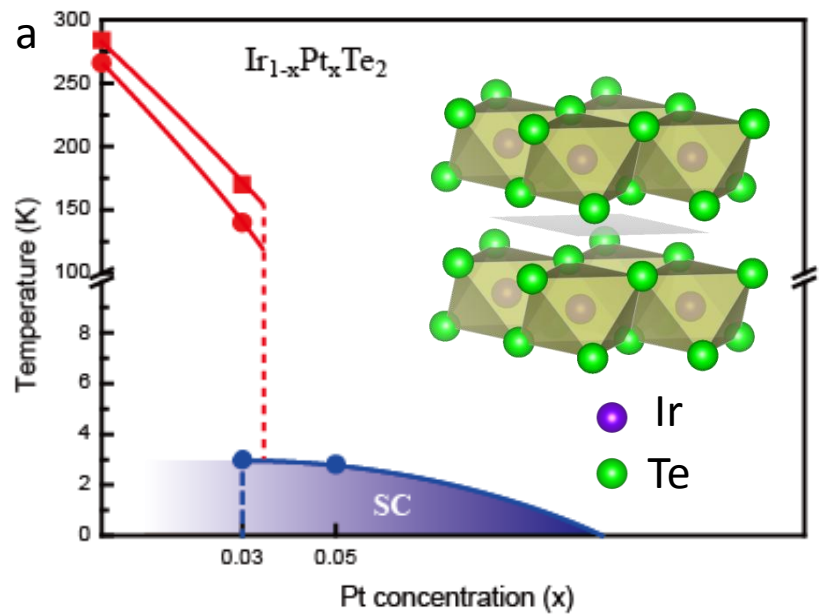
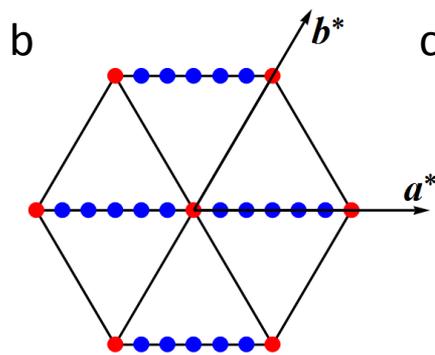
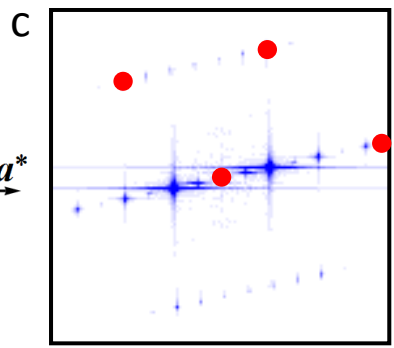
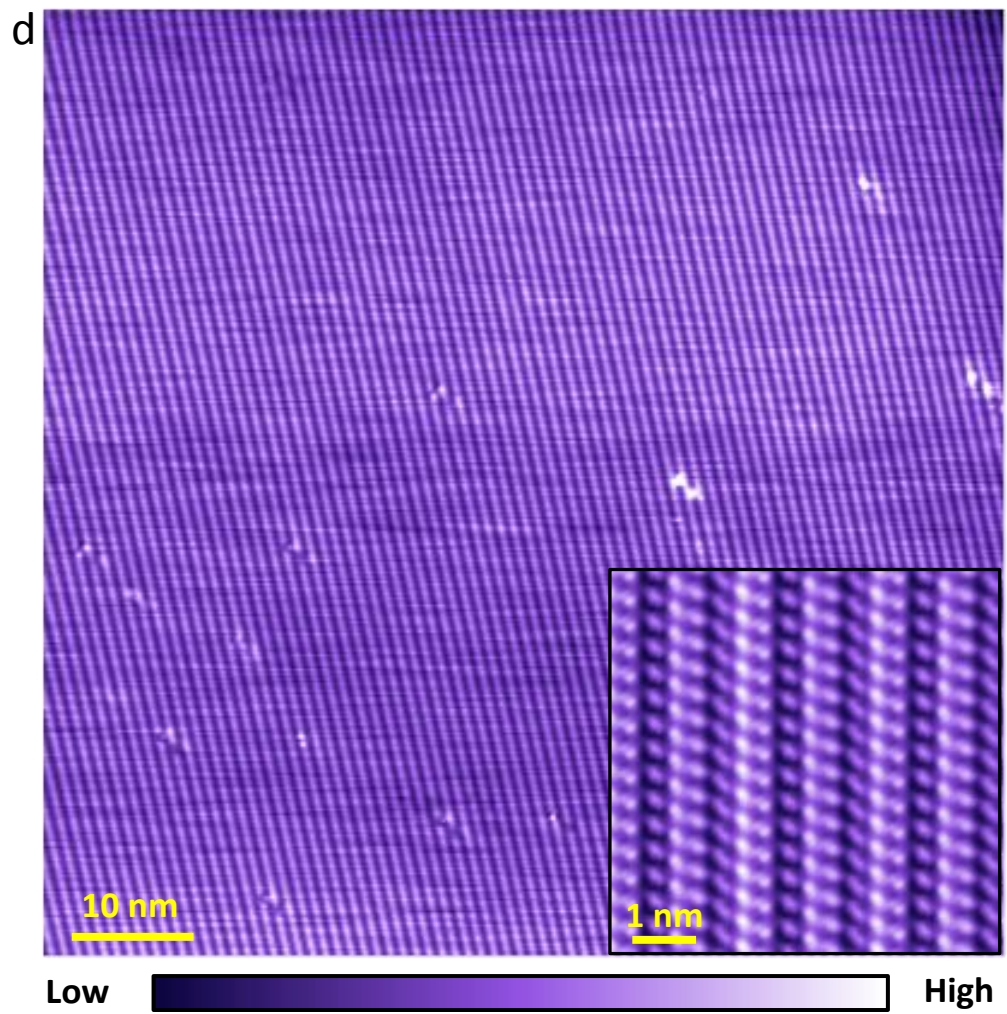

Figure 1

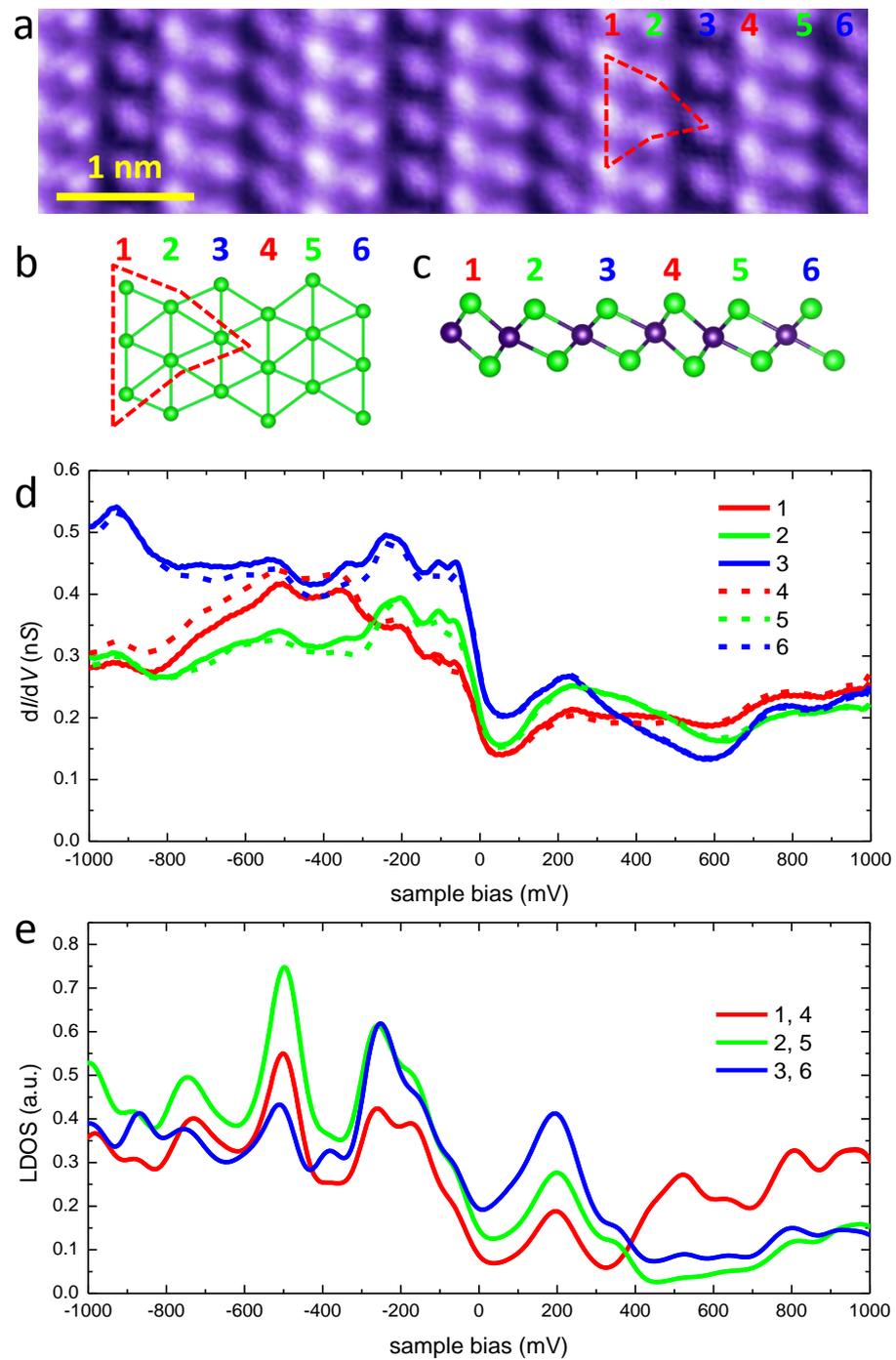

Figure 2

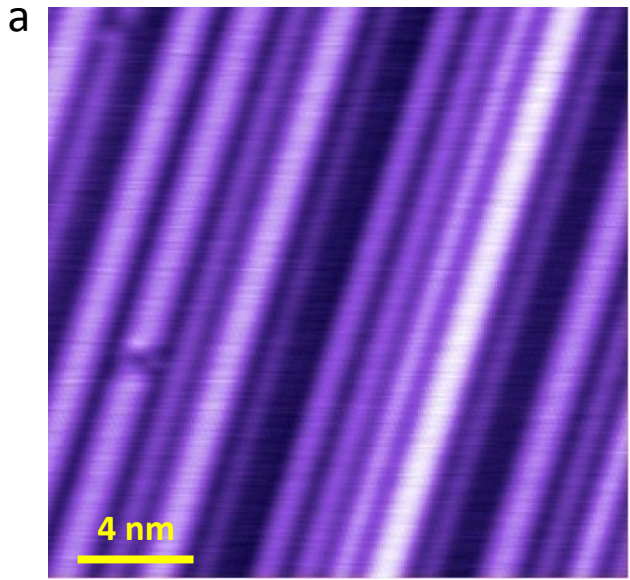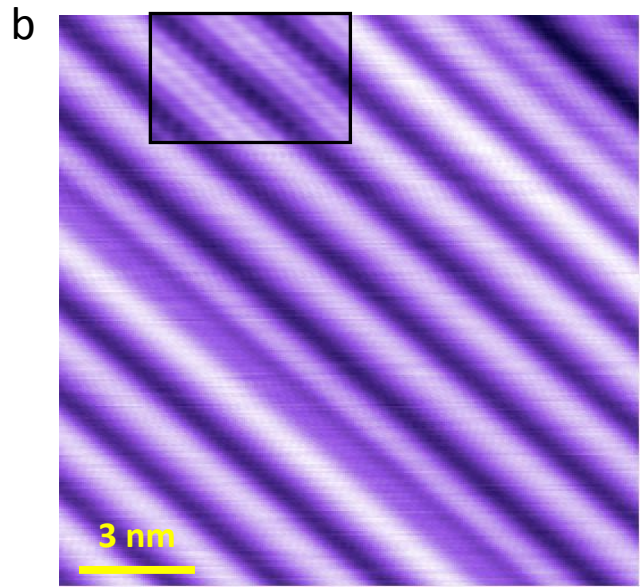

Figure 3

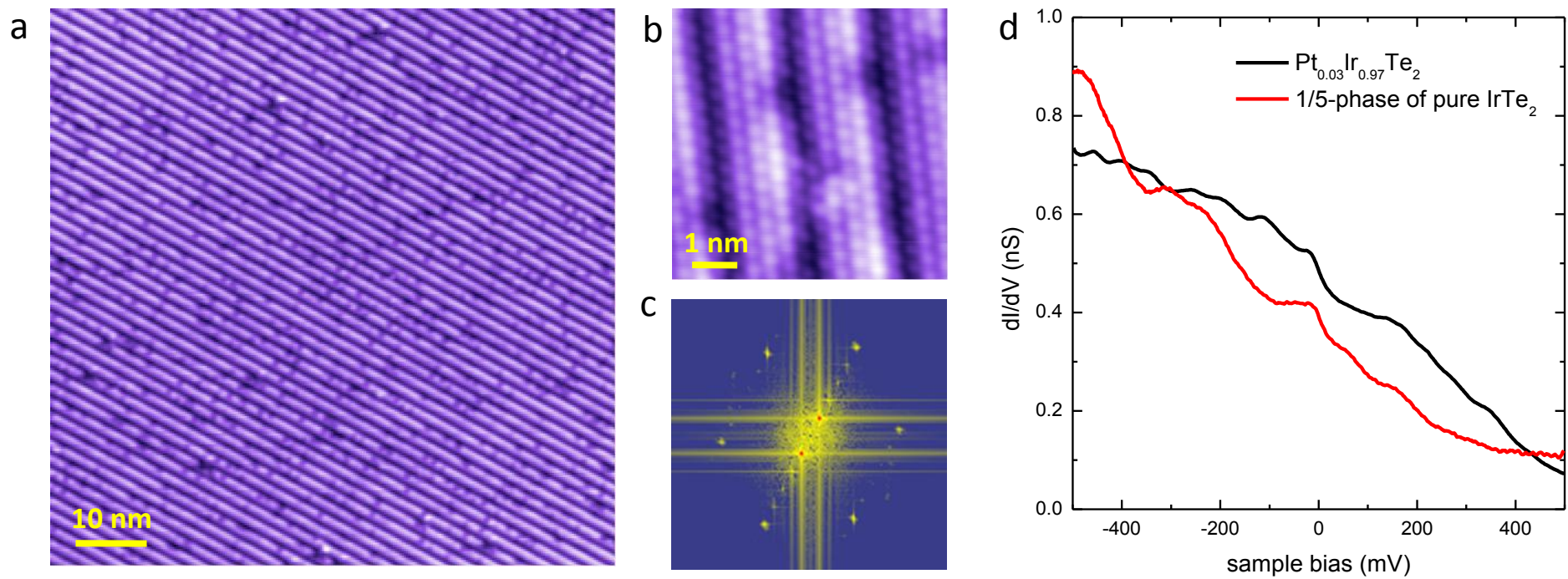

Figure 4

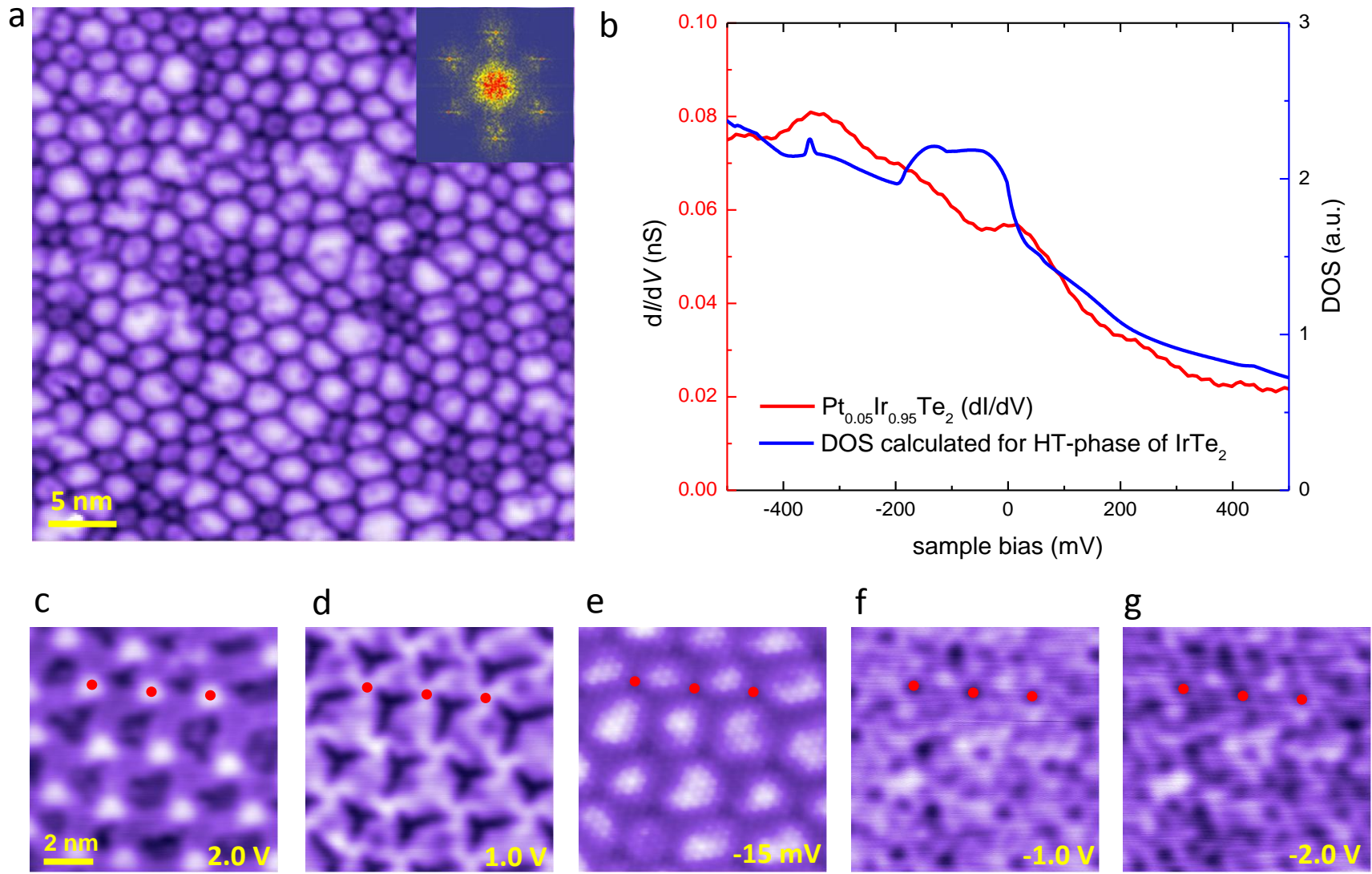

Figure 5

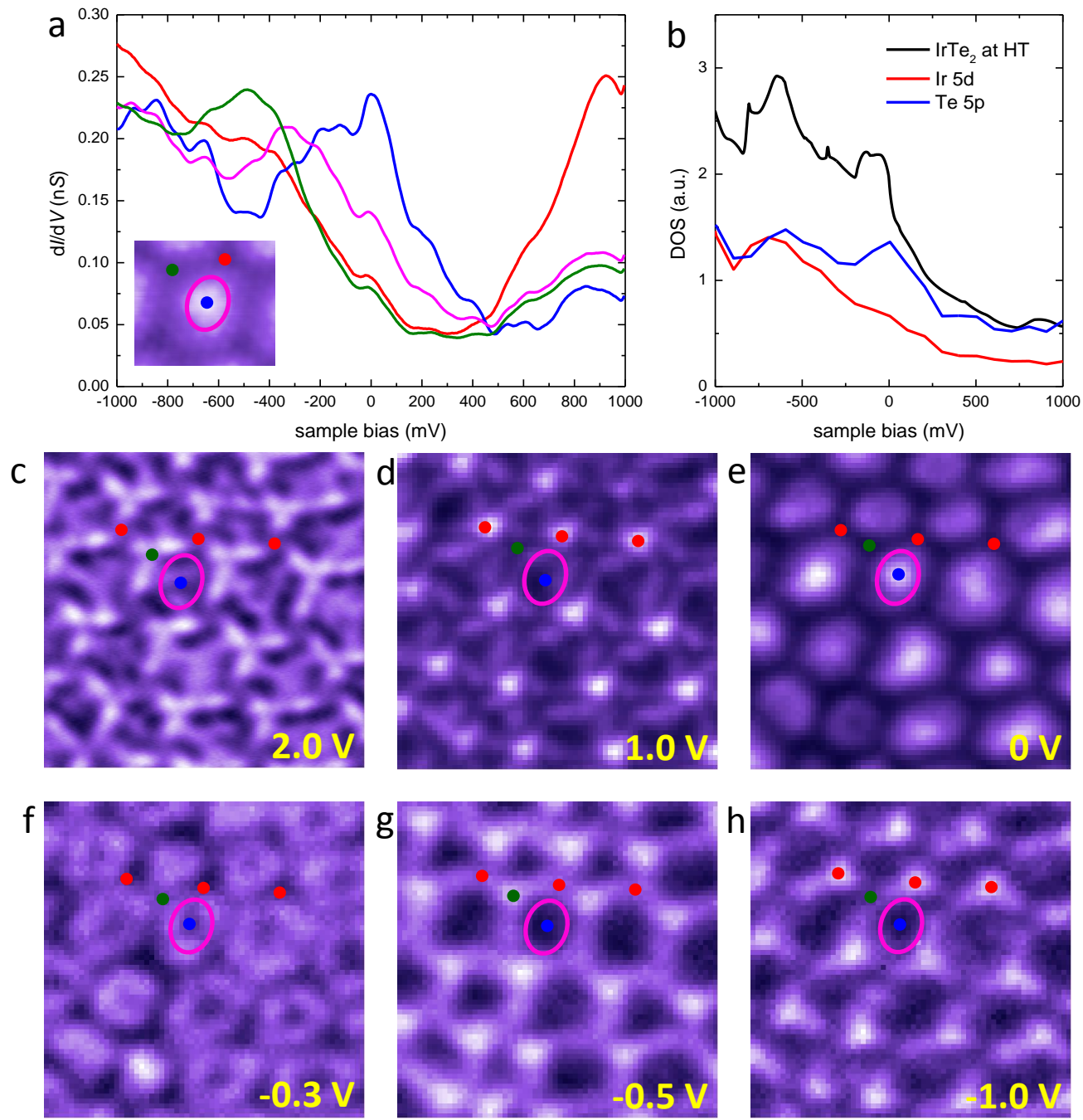

Figure 6